# Evaluating Serverless Machine Learning Performance on Google Cloud Run

1st Prerana Khatiwada
Computer and Information Sciences
University of Delaware
Newark, United States
preranak@udel.edu

2nd Pranjal Dhakal
Computer and Information Sciences
University of Delaware
Newark, United States
dpranjal@udel.edu

*Abstract*—End-users can get functions-as-a-Service from serverless platforms, which promise lower hosting costs, high availability, fault tolerance, and dynamic flexibility for hosting individual functions known as microservices. Machine learning tools are seen to be reliably useful, and the services created using these tools are in increasing demand on a large scale. The serverless platforms are uniquely suited for hosting these machine learning services to be used for large-scale applications. These platforms are well known for their cost efficiency, fault tolerance, resource scaling, robust APIs for communication, and global reach. However, machine learning services are different from the web-services in that these serverless platforms were originally designed to host web services. We aimed to understand how these serverless platforms handle machine learning workloads with our study. We examine machine learning performance on one of the serverless platforms - Google Cloud Run which is a GPU-less infrastructure that is not designed for machine learning application deployment.

## I. INTRODUCTION

Serverless computing platforms eliminate and minimize the need for expensive onsite hardware, software, and storage infrastructure. These platforms run containerized applications, which are very cost-effective. Therefore, it's important to know how to package such applications as containers and to be familiar with some of the technical terms we'll come across while working on this project. Through this project, we aim to understand microservice architecture, which is an increasingly popular approach to software development. The below section provides a more detailed description of the terms that are commonly used for this project and are discussed further in the rest of the section.

Machine learning applications are commonly deployed in servers with GPUs, which are very costly. These GPU-enabled platforms offer a limited set of services compared to serverless platforms like Google Cloud Platform or Amazon Web Services. Thus, the goal of this study is to understand how machine learning web applications perform in serverless platforms that aren't yet optimized for this sort of workload.

### A. Containers

A container is an executable entity consisting of software code along with the operating system libraries and dependen-cies required to run the code and it enables applications to run almost anywhere. The process of creating containers is called containerization [1]. Containers are portable and are extensively used for developing modern cloud applications. Containers virtualize the operating system, allowing them to run anywhere from a private data center to the public cloud, or even on a developer's laptop. Everything at Google runs in containers, from Gmail to YouTube to Search [2].

Containers and virtual machines are comparable in that they promote IT efficiency and DevOps. However understanding difference between them is critical to build a cloud native approach. A virtual machine has a guest operating system, a virtual copy of the hardware that the operating system needs to run, and an application with its associated libraries and depen-dencies [3]. Unlike a virtual machine, containers do not need to include the guest operating system in every single instance, instead it can leverage the features and resources of the host operating system so that each individual container contains only the applications and its libraries and dependencies.

### B. Docker

The host machine (server) where the containers are ex-ecuted has an open-source runtime engine like the Docker runtime engine installed on its operating system [4]. The host's operating system services are shared by the containers. Docker uses resource isolation in the OS kernel to run multiple containers on the same OS, and these containers are well-suited to microservices, which is the traditional approach known as the monolith architecture of software development that involves developing solutions as one large system [5]. The components in the monolith architecture are tightly coupled. So, making minor changes is a very involved and lengthy process. The microservices architecture, in contrast, divides the application into smaller functional units, making them loosely coupled. These functional units or services are devel-oped, containerized, and deployed independently. If one of the services requires scaling, it can easily be done without scaling the entire application, leading to efficient resource utilization.

### C. REST API

Microservices communicate with each other using REST Application Programming Interfaces (APIs) [6]. REST stands

for Representational State Transfer. It defines a set of rules/constraints for a web application to send and receive data [7]. REST APIs use HTTP requests and can be developed using any programming language. For this work, we have used the python rest API to demonstrate how we can easily dockerize a Flask App. It is a simple API that gives an image URL it returns its category. We take this API, put it in a docker container, and test it out using Postman.

### D. Google Cloud Run

To develop and deploy highly scalable containerized appli-cations on a serverless platform [8], we will use Google Cloud Run, a service provided by Google. Developers can build and run applications without having to manage servers. It performs automatic scaling depending on the request traffic. During the idle period, it even scales down to 0 and only charges for the resources we have used. It also provides logging and monitoring features.

### E. Flask Application Programming Interface

A web application framework is a collection of libraries and modules that takes care of low-level details such as protocol and thread management [9]. Flask is a web appli-cation framework developed by Armin Ronacher and written in Python programming language to help developers write web applications quickly. It is often called Flask microframework and is designed to keep the core of the application simple and scalable. It offers URL routing and template engine features.

### F. Image Classification

Image classification is a way of categorizing the input images into pre-defined labels or categories [10]. The clas-sification models learn from the images dataset that we train, allowing us to make predictions in the end. In our work, we have created a machine learning Python Flask application for image classification. We have used a pre-trained MobileNetV2 Tensorflow Model to classify images into one of the 1000 categories which is based on the ImageNet benchmark dataset. This classifier does not require initial training [11]. Pretrained models are deep neural networks that are trained using a large images dataset [12].

## II. RELATED WORK

"Cloud services" are gradually becoming an important com-ponent for businesses of all sizes [13]. This allows them to store data on a third-party cloud service like Microsoft Azure or Amazon Web Services. Microservices are a recent concept in the software development world. James Lewis first proposed the notion at a conference in San Francisco in 2012 [14]. Netflix was one of the first companies to use the concept on a big scale, referring to it as "fine-grained service-oriented architecture" [15]. Microservices are built on these earlier techniques, and some argue that microservices and service-oriented architecture are interchangeable terms. On the other hand, microservices are smaller and do not share data storage [16].

## III. METHODOLOGY

To accomplish our goal of evaluating the machine learning performance in the Google cloud platform, we start with the engineering task of developing a machine learning web appli-cation and deploying it in the Google Cloud Run Platform. The system overview is shown in the figure 1.

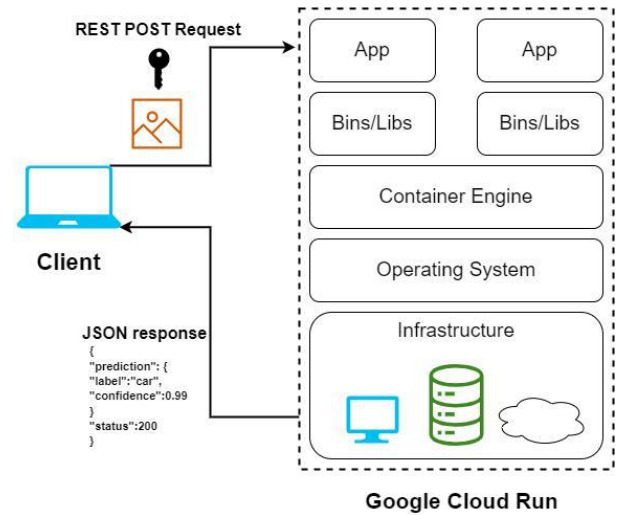

Fig. 1. System Diagram.

### A. Creating an API

The web application we have developed provides REST APIs that the client devices can use to send requests with proper authentication. We have written 2 endpoints in our web application. The first one is a health check GET endpoint /healthz that returns a success text string. The second is a POST endpoint /predict. This app authenticates the POST request and, if valid, accepts images from the user. The app runs a TensorFlow MobileNetV2 Image Classification model [17]. This model can classify images into one of the 1000 categories.

### B. Containerizing the application

Once the web application is built, we containerize the application using docker. We have specified to use python:3.8-slim which is a paired down version of the full Debian image and only has minimal packages to run Python [18]. A docker image is then built using the Dockerfile. This image is pushed to the Google Container Registry service which can be picked by the Google Cloud Run for deployment.

### C. Deployment in Google Cloud Run

The containerized image is then deployed on Google's Cloud Run platform. We set the container port to 5000 as the Flask app uses this port for communication. Memory and CPU capacity are set to be 2GiB and 1 respectively as we determined this was the minimal configuration required to run our application. After the deployment, a public URL is gener-ated that can be used for health check requests and prediction requests. Deploying a Flask application with Docker will allow us to replicate the application across different servers with minimal reconfiguration. In this project, we developed and deployed highly scalable containerized applications on a fully managed serverless platform.

To simulate different load scenarios and stress test the server, we use the Locust API [19]. Locust is a scriptable and scalable performance testing tool written in Python. We generate metrics like response latency, cold start time, memory and CPU utilization, and crash frequency. We learn how the Cloud Run platform performs load balancing and handles scaling.

## IV. EVALUATION

The evaluation is divided into two parts: client-side evalua-tion and server-side evaluation. On the client-side, we simulate a situation in which multiple requests are sent to the server in parallel. First, we generate metrics that will be of interest to the customer. The following are the metrics:

- Request Statistics - We keep track of the endpoint and how many times a request is sent to it. We also monitor the number of failures, the average time it takes for the request to finish, and the size of the object requested.
- Response Time Statistics - This is concerned with the transport latency and the processing time in the server.
- Number of Users - Multiple users can make requests to the server concurrently. The number of users at any given time is measured here.

We use Cloud Run's logging functionality to do the server-side evaluation. Cloud Run log files provide a wealth of information about each request and the state of the container, which can be parsed to generate a variety of metrics. The following metrics are computed on the server-side.

- Container Instance Count - Cloud Run scales resources automatically. Cloud Run may spin-off additional con-tainers to accommodate requests or terminate containers for more efficient resource use, depending on the number of requests arriving at the server. We keep track of the number of containers against time.
- Billable Container Instance Time - Cloud Run only bills when the server is processing requests. We can validate this claim by plotting the billed time against time.
- Request Count - We use the server-side request count and compare it to the client-side count to see if they match.
- Request Latency - Measures the request processing time in the server.
- Container CPU Utilization - Measures the CPU utilization of the containers.
- Container Memory Utilization - Measures the memory utilization of the containers.

### A. Client-side Evaluation

From the client-side side evaluation, there are some interest-ing observations to be made. We can start with some obvious findings, such as the fact that the GET request uses fewer resources and has a shorter response time. All 425 requests were completed successfully and returned with a status code of 200. Even when the number of requests per second was drastically increased, the server did not drop any of them. We can see in figure 2 that initially the response time is quite high of almost 20 seconds and then it reduces to almost 2 seconds averaging at 10 seconds. If we continue to send additional requests, this average will continue to fall. The cold-start problem is to account for this response time behavior. The Cloud Run platform terminates all the containers during sustained idle time periods to be cost-efficient. As a result, new containers are created as new requests arrive. This increases the response time in the beginning. Cloud Run, on the other hand, offers the option of maintaining some containers running even during idle time to address this issue at the cost of extra billable hours.

TABLE I
CLIENT-SIDE REQUEST STATISTICS

| Method | GET | POST | |
|---|---|---|---|
| Endpoint Name | / | /predict | Aggregated |
| Number of Requests | 90 | 335 | 425 |
| Number of Failures | 0 | 0 | 0 |
| Average (ms) | 770 | 1198 | 1107 |
| Min (ms) | 104 | 132 | 104 |
| Max (ms) | 18287 | 21187 | 21187 |
| Average size (bytes) | 48 | 48 | 48 |
| Requests Per Second | 0.9 | 3.5 | 4.4 |
| Failures per seconds | 0 | 0 | 0 |

TABLE II
CLIENT-SIDE RESPONSE TIME STATISTICS

| Method | GET | POST | |
|---|---|---|---|
| Endpoint Name | / | /predict | Aggregated |
| 50%ile (ms) | 580 | 700 | 680 |
| 60%ile (ms) | 600 | 790 | 720 |
| 70%ile (ms) | 680 | 900 | 810 |
| 80%ile (ms) | 770 | 1000 | 1000 |
| 90%ile (ms) | 990 | 1100 | 1100 |
| 95%ile (ms) | 1100 | 1300 | 1300 |
| 99%ile (ms) | 18000 | 18000 | 18000 |
| 100%ile (ms) | 18000 | 21000 | 21000 |

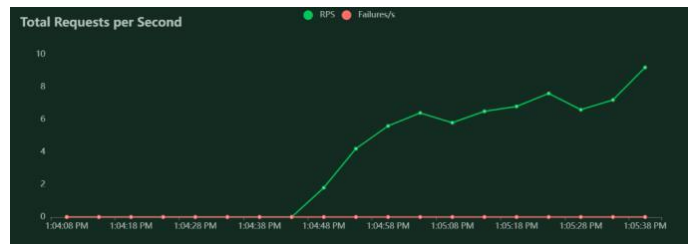

Fig. 2. Client-side: Total Requests per second.

### B. Server-side Evaluation

In figure 5, we see that Cloud Run automatically performs resource scaling and increases the number of containers to accommodate all requests coming to the server. Once the requests stop arriving, the number of idle containers increase, and active ones decrease. When containers are idle for some period, they are terminated for cost efficiency. The user is only billed for time-period in which the containers were being utilized which can be verified from figure
6. The request latency in figure 8 has a similar graph to graph in figure 3. This is the cold-start problem in effect described earlier. The CPU utilization of containers in figure 9 shows the utilization is at its peak when we are actively sending requests. It is interesting to see that the memory utilization in 10 remains quite high during the entire time the container exists even when there are no active requests. This is due to the fact that the containers load entire operating system and necessary dependencies into its RAM to process the requests.

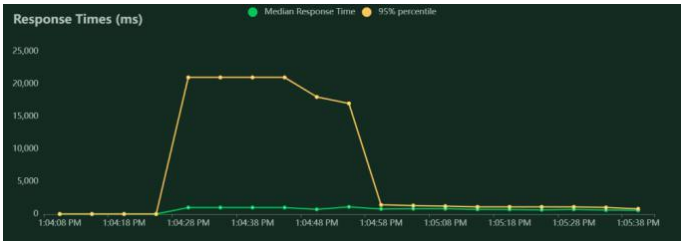

Fig. 3. Client-side: Response time (ms).

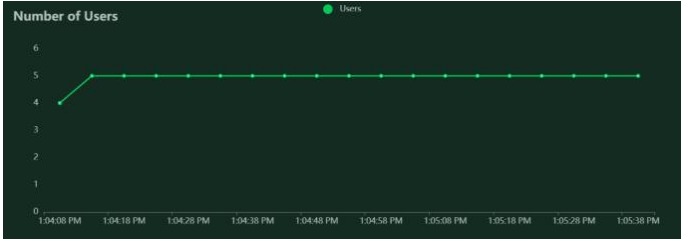

Fig. 4. Client-side: Number of Users.

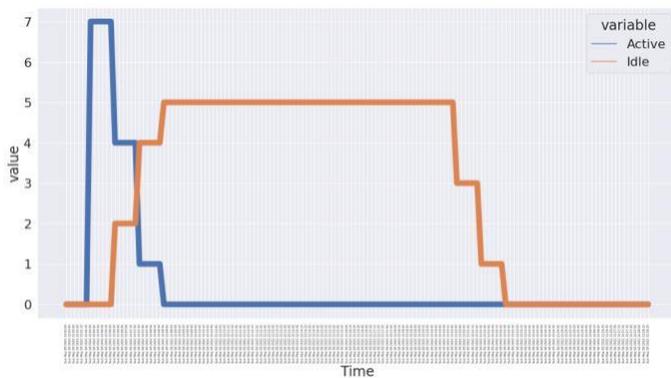

Fig. 5. Server-side: Container Instances vs Time.

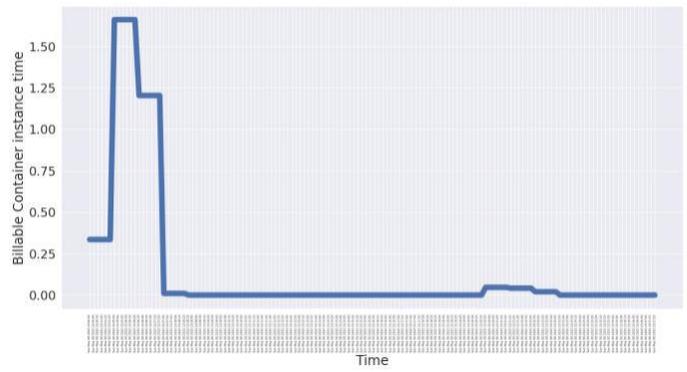

Fig. 6. Server-side: Billable Container Instance Time vs Time.

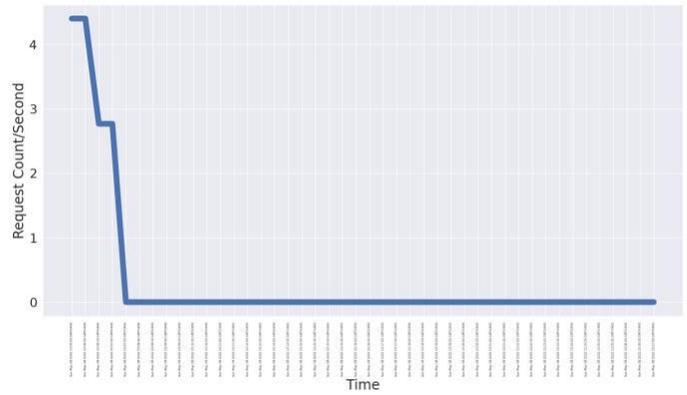

Fig. 7. Server-side: Request Count vs Time.

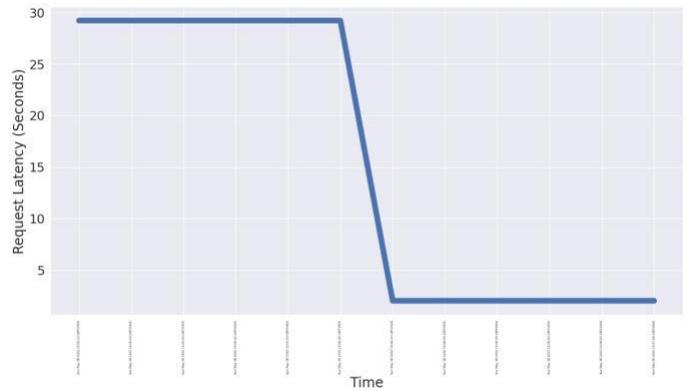

Fig. 8. Server-side: Request Latency vs Time.

performance using server and client-side statistics. We also observed that machine learning applications can be deployed on this platform with performance comparable to standard web apps, which this platform was built to host. Even under high test load, the response time was good, and no requests were dropped.

## V. CONCLUSION

We created a web application that runs an image clas-sification model for this project, containerized the web ap-plication, deployed it on Google Cloud Run, and analyzed

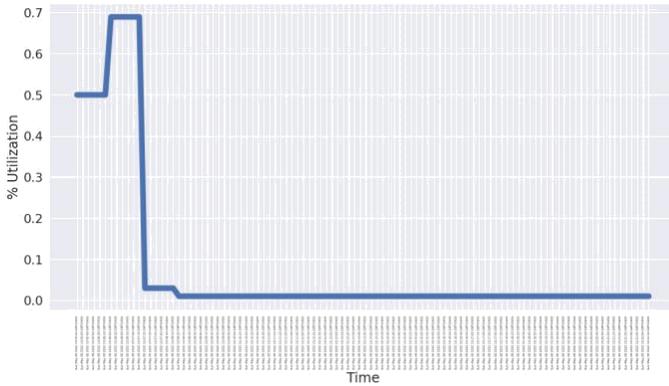

Fig. 9. Server-side: CPU Utilization vs Time.

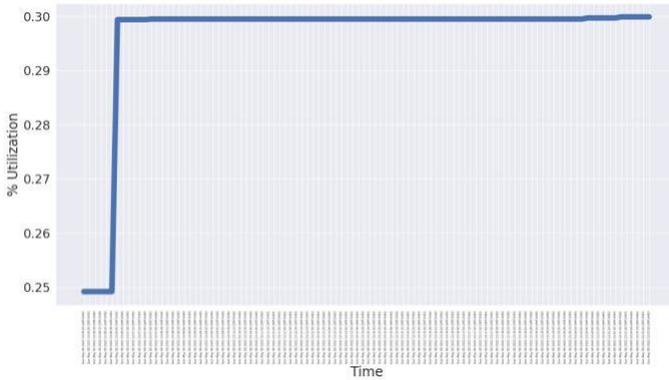

Fig. 10. Server-side: Memory Utilization vs Time.

## VII. ACKNOWLEDGMENT

We would like to express our gratitude and appreciation to all those who gave us the opportunity to complete this work. Special thanks are due to our Professor Prof.Lena Mashayekhy whose help, encouragement, and constant supervision helped us in all-time in selecting a suitable area to work and in writing this manuscript. Our profound thanks go to all our peers, especially to friends for spending their time helping, reviewing our work and giving us support whenever we need it in improving our project idea. We would like to extend our sincere thanks to all of them.